# An Instrument-Free Demonstration of Quantum Key Distribution for High-School Students

María José Carreño[b], Jonathan Sepúlveda[a], Silvia Tecpan[a], Carla Hernández[a], Felipe Herrera[a,b]

[a]Department of Physics, Universidad de Santiago de Chile, Av. Ecuador 3493, Santiago.
[b]Millennium Institute for Research in Optics, Concepción.

It has become increasingly common for high-school students to see media reports on the importance of quantum mechanics in the development of next-generation industries such as drug development and secure communication, but few of them have been exposed to fundamental quantum mechanical concepts in a meaningful classroom activity. In order to bridge this gap, we design and test a low-cost 20-minute demonstration of the Bell test, which is used in several entanglement-based quantum key distribution protocols. The demonstration introduces ideas such as the quantum state, quantum measurement, spin quantization, cryptography, and entanglement; all without using concepts beyond the 9*th* grade of the Chilean high-school curriculum. The demonstration can serve to promote early exposure of the future adopters and developers of quantum technology with its conceptual building blocks, and also to educate the general public about the importance of quantum mechanics in modern industry.

## Introduction

Quantum mechanics is the most precise physical theory that mankind has developed to date [1], yet learning concepts such as spin quantization or quantum entanglement is usually associated with higher level university training, and within a restricted set of disciplines. At the high-school level, advanced quantum mechanical concepts are practically absent in most study programs [2]. In the wake of the so-called *second quantum revolution* [3], whose outcomes are likely to be enjoyed by the public in only a few decades, it becomes necessary to start introducing the fundamental concepts of quantum mechanics at the earliest stages of STEM education (Science, Technology, Engineering and Mathematics). This can be achieved by developing teaching material that can help school teachers prepare meaningful classroom activities, without expensive materials or equipment. It could be expected that conceptually accessible teaching material can promote early exposure to quantum science for a generation who will likely become the early adopters and developers of mass-scale quantum technology. It may also serve as a tool for educating the general public about the importance of quantum mechanics in modern industry.

In an effort to bridge a growing gap between modern quantum science and high school education, in this work we develop a low-cost demonstration of the Bell test. The demonstration highlights some of the key aspects of quantum entanglement between particles. In a Bell test, two communicating parties (Alice and Bob) analyze the statistical correlations between measurements they each carry out on individual particles that are possibly entangled [4]. The question of whether the observed statistical correlations can be accounted for using a classical or a quantum description of reality is settled by the so-called *Bell inequality* [4]. Alice and Bob need to communicate each other the results of their individual particle measurements, using a classical communication channel (phone, radio, internet). By comparing their results, they compute a single number that depends on the statistical correlations between their measurements. Whenever this number exceeds a previously defined threshold ( $2\sqrt{2}$ for two entangled qubits [4]), Alice and Bob can be sure that the particles they each received were quantum entangled; otherwise, they were not.

Nowadays, there are secure communication protocols that involve some type of Bell test to verify quantum entanglement [5]. Consider optical communications for example: one encodes messages (binary strings) in the internal states of light quanta (photons), such as the polarization state, then transmit encoded light through free space or optical fibers between two distant locations. Clearly this has been done for decades and no quantum entanglement is needed for optical communication to work. The novelty of quantum-secure optical communication, also known as optical quantum cryptography [6], is the ability to guarantee the security of the optical message by detecting whether two light sources are entangled or not. Why would entanglement between flying photons (or spins) guarantee the security of a message? This is the key conceptual insight that our demonstration aims to provide to high-school teachers and students.

## Bell Test Demonstration

**Conceptual Preparations**

Prior to the demonstration, instructors should discuss three key concepts: *encryption*, *quantum bit*, and *spin quantization*. For the first concept, instructors can motivate the definition of a *cryptographic key,* used in secure communication for encrypting and decrypting a message, such that only emitter and receiver have access to the key (shared key), and any unauthorized party (eavesdropper) cannot.

In order to introduce the idea of a quantum bit (*qubit*), comparisons with classical bits in electronics can be made, since students are likely to be familiar with digital electronics. One possible comparison can exploit the restricted number of possible configurations that a classical bit can take (*on* = 1, *off* = 0), versus the in principle infinite number of configurations in which a qubit can be prepared. We argue that the Bloch sphere picture of the qubit [1] can be intuitive enough to illustrate this concept, without elaborating on the underlying complex calculus. Instructors can either draw on a whiteboard or show an image of a globe where the north pole corresponds to one classical bit value (*on*) and the south pole to the opposite bit value (*off*). Then explain that any point on the surface of the globe represents one possible state of the qubit. This gives students a sense for the extended range of outcome possibilities that the qubit comparatively offers.

To further elaborate on the qubit concept, instructors can discuss how a quantum bit can be implemented in the laboratory. In other words, the qubit is not only a mathematical concept (Bloch sphere), but is also a tangible physical object that can be engineered. One possible physical system to illustrate this point could be the spin of the unpaired electron in the outer shell (5*s*) of a silver atom. The electron spin can be introduced here as two alternative ways in which the electron can rotate about itself. Each rotation direction (spin up and down) defines the north and south poles of a Bloch sphere. Additional examples from the literature could be used to illustrate the *quantization of spin* [1]. The demonstration we describe below builds on the idea that a quantum particle can have an internal property (spin) that upon measurement gives two possible outcomes (spin up or down) that can be recorded on a data table.

**Materials**

- 2 non-transparent bowls with 25-30 cm diameter.
- 4 identical colored foam spheres with 5 cm diameter (2 red, 2 blue).
- 2 pieces of blindfold cloth.
- [*Optional*] 3 white poster paper sheets 76 cm x 120 cm.
- [*Optional*] 40 pieces of square color paper 10 cm x 10 cm (20 blue, 20 red).
- [*Optional*] 20 pieces of black color paper (10 triangles, 10 circles).

The two bowls must not be transparent and each should be clearly labelled with the numbers "1" and "2". In our test implementation we used fish bowls, as Fig. 1 shows. The bowls could in principle be replaced with any non-transparent container in which it is comfortable to place the red and the blue foam spheres, such that every time a blindsided volunteer picks one sphere from the container, she or he has equal chances of picking either color. We list paper material above as *optional* because everything that is planned to be done with poster papers and colored cards, can also be done on a whiteboard with colored markers.

**Description and Interpretation of the Demo**

The demonstration requires four volunteers and one instructor to be successfully implemented. The role of each person involved is as follows:

- **Volunteer A**: Represents an experimentalist (*Alice*).
- **Volunteer B**: Represents a second experimentalist (*Bob*).
- **Volunteer C**: Represents the particle source, particle delivery and simulates particle entanglement.
- **Volunteer D**: Registers the measurement outcomes obtained by Volunteers A and B, as well as the results of the joint measurement outcomes.
- **Instructor:** Narrates the experiments carried out by Volunteers A and B, analyzes the joint measurement outcomes, and interacts with the audience.

The **Instructor** narrates the steps of the demonstration using supporting slides [8], and explains important concepts that emerge as the demonstration develops. **Volunteers C and D** have a key practical role in the successful implementation of the demonstration and should be involved in its preparation as much as possible. **Volunteers A and B** can be students selected from the audience, or anyone with preferably no knowledge about the specifics of the demonstration.

Before the demonstration begins, **Volunteer C** puts one red sphere and one blue sphere inside the *Particle 1* and *Particle 2* containers. Then place both filled particle containers on top of a small table facing the audience. The number labels should be visible to the audience at all times for clarity (see test example in Fig. 1). Because these preparations are done in front of the audience, students know that there are two spheres of different colors (red and blue) inside each particle container. The **Instructor** then explains that each container represents a quantum particle and that the spheres of different colors inside them represent the two possible values for the particle spin, which is an internal property of the particle that must be measured to find out its value. The **Instructor** also explains that in this simulated experiment, the table on which the containers are placed represents the source of a pair of particles. Without explaining what the word entanglement means, the **Instructor** anticipates at this point that two sets of experiments will be carried out sequentially: one first set in which the two particles that come out of the source (table) *are not* entangled, and a second set in which the particles *are* entangled at the source. The audience is then encouraged to follow each step of the demonstration carefully to try to *find out* what is the practical consequence of entanglement in the second set of experiments.

The **Instructor** then asks two students from the audience to become **Volunteers A and B**. **Volunteer A** is referred to as *Alice* and **Volunteer B** as *Bob* throughout the demonstration. Alice and Bob are then placed on either side of the particle source (table), facing the audience. **Volunteer C** then proceeds to cover the eyes of **Volunteers A and B** with blindfolds, making sure that they are unable to see through. This detail is important because it simulates the lack of communication between Alice to Bob when measurements are carried out on the individual particles they receive.

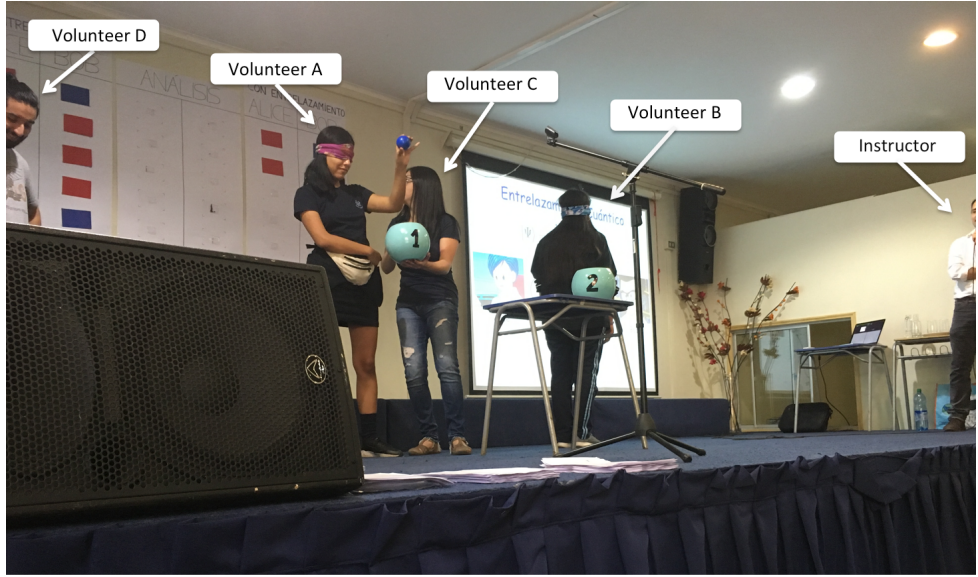

**Figure 1**. Illustration of the setup including four volunteers, one instructor, and AV support. In the picture, Volunteer A is receiving *Particle 1* (labelled container) from Volunteer C and measures its *spin projection* (picks a colored sphere). Volunteer D is recording the measurement outcome on a data table (poster paper), while Volunteer B waits to receive *Particle 2* (container) from Volunteer C. The Instructor narrates all the steps to the audience.

The set of experiments *without* entanglement between the particles begins when **Volunteer C** takes *Particle 1* (bowl) from the particle source (table) and allows **Volunteer A** (Alice) to perform a measurement of the particle spin projection (pick a colored sphere). The **Instructor** narrates these steps and asks **Volunteer A** to show the result of the measurement outcome (color) to the audience, as shown in Fig. 1. **Volunteer D** registers the measurement outcome on a two-column data table that has the title "*Without Entanglement*" on top. In one column of the data table, the measurement outcomes of **Volunteer A** are registered, and those of **Volunteer B** are recorded on the other column. After Alice's first measurement outcome is registered in the appropriate entry of the data table, **Volunteer C** puts the *Particle 1* container, with its two colored spheres inside, back on top of the source (table). Then **Volunteer C** proceeds to take the *Particle 2* container and let **Volunteer B** (Bob) perform his first measurement of the *Particle 2* spin projection (pick a sphere). The measurement outcome (color) is again recorded by **Volunteer D** on the appropriate entry of the data table. The first repetition of this experiment concludes when **Volunteer C** puts the *Particle 2* container, with its two colored spheres inside, back on the source. The **Instructor** verbally emphasizes that this step concludes the first experimental run.

Each experimental run is then repeated at least five times. This is done for two main reasons:

- To simulate the fact that in a quantum world, particles cannot be associated with a predefined value of a physical observable (spin projection) until we carry out a measurement of that observable and record the outcome. The same particle can *in general* lead to different measurement outcomes for the same observable if measured repeatedly.
- To reduce the probability that the two data tables built to record the outcomes of the experiments *with* and *without* entanglement become indistinguishable from each other after the second set of experiments is finished. This simulates the closing of the so-called measurement loophole in Bell tests [4,9], which roughly states that an entanglement test via Bell inequality cannot be considered conclusive if not enough statistics is collected to carry out a definite correlation analysis.

After all repetitions of the particle pair measurements by **Volunteers A and B** are done, the audience will see the data table entitled "*Without Entanglement*" filled with all five pairs of outcomes obtained by Alice and Bob, as shown in Fig. 2a (leftmost data table).

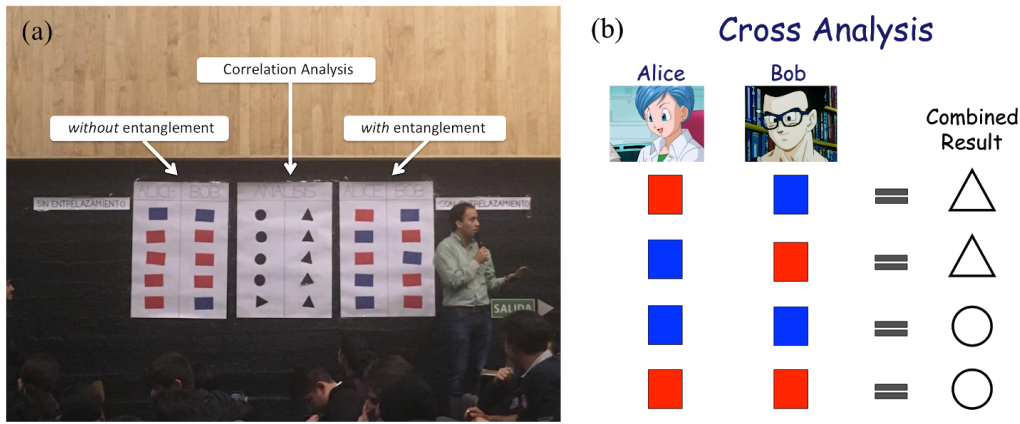

**Figure 2**. (a) Real example of measurement records for the set of experiments with entanglement (rightmost data table) and without entanglement (leftmost data table) between each pair of particles sent to Alice and Bob. The middle data table is the correlation analysis of the joint measurement outcomes. (b) Slide used to define the symbol coding for the simulated cross correlation analysis ("Cross Analysis") of the measurement outcomes obtained independently by Alice and Bob, for each pair of particles they received from Volunteer C. If Alice and Bob measure different spin projections (sphere colors) in their experiments, a triangle is assigned to the joint outcome ("Combined Result"). Circles are assigned to the joint outcome, when the same spin projections were measured.

Before the second set of experiments begins, the **Instructor** explains that now the two particles emitted by the source will now be prepared in a *quantum state* that can be written as the sum

$$\Psi = (\uparrow)_1 (\downarrow)_2 + (\downarrow)_1 (\uparrow)_2 \qquad (1)$$

where the up arrow represents spin up (red color) and the blue arrow represents spin down (blue color). This expression for the two-particle quantum state can be either drawn on a whiteboard, or put on a supporting slide.

The **Instructor** explains that the quantum state of the two particles must be represented by such a summation, or *coherent superposition*, of two equivalent scenarios. The first term in the superposition represents the case where *Particle 1* comes out of the source with its spin down (red) and *Particle 2* comes out with its spin down (blue). The second term represents the case where the particles come out of the source with the opposite combination of colors. Both possibilities are equally likely to occur because there is a physical process happening at the source that makes these two scenarios (pathways) indistinguishable. This introduces the notion that in quantum mechanics, indistinguishability of pathways leads to coherent superpositions. The **Instructor** explains that the coherent superposition in Eq. (1) represents an entangled joint state of *Particles 1* and *2*.

The set of experiments *with* entanglement begins in the same way as the first set of experiments: **Volunteer A** measures the spin projection (picks a sphere) of *Particle 1*, **Volunteer D** records the measurement outcome (color) on the appropriate column of the data table entitled "*With Entanglement*", and **Volunteer C** places the *Particle 1* container with the two colored spheres back on the particle source (table). At this point, **Volunteer C** must enforce the behavior expected for the entangled two-particle state in Eq. (1). This is done as follows:

- If **Volunteer A** picks the red sphere (spin up) upon measuring *Particle 1,* then **Volunteer C** withdraws the red sphere from the *Particle 2* container.
- If **Volunteer A** picks the blue sphere (spin down) upon measuring *Particle 1,* then **Volunteer C** withdraws the blue sphere from the *Particle 2* container.

By doing this, **Volunteer C** is effectively simulating a *projective measurement* on the entangled two-particle state represented by Eq. (1). The measurement done by **Volunteer A** on *Particle 1* is said to *project* the state of *Particle 2* in a way that is consistent with the structure of the entangled state in which the particle pair was prepared. This in turn reduces the possible outcomes that **Volunteer B** is allowed to obtain. Therefore, once **Volunteer B** receives the *Particle 2* container having only one sphere inside, the only possible outcome (sphere color) that can be recorded is the one enforced by the simulated projective measurement. Note that because they are blindfolded, **Volunteers A** and **B** are unaware of what **Volunteer C** did to simulate entanglement in the *Particle 2* container. The **Instructor** also asks the audience beforehand to remain silent during the experimental run. This simulates the fact that both Alice and Bob are only aware of their own individual particles. Each of them does not necessarily know that there is a second particle being measured somewhere else by a second party, nor that the two particles involved may be entangled. The **Instructor** can emphasize this by saying that Alice and Bob can in principle live in two separate galaxies. All they each do is to receive a quantum particle from a source, measure its spin projection and record the outcome on a data table, in order to learn something about the particle's quantum state. They only become aware of the second party and her or his measurement outcomes at a later stage, when a correlation analysis is carried out.

The experimental run *with* entanglement is then repeated the same number of times as its was done in the experiments *without* entanglement. The results of the spin measurements carried out by Alice and Bob for each repetition are recorded by **Volunteer D** as in Fig. 2a (rightmost data table).

Once the experiments *with* and *without* entanglement are done, **Volunteer C** removes the blindfolds from **Volunteers A and B,** and they return to their seats. The **Instructor** then introduces the idea that in order to quantitatively assess the presence or absence of entanglement between the particles, Alice and Bob must carry out a cross correlation analysis. The **Instructor** explains that in order to perform such correlation analysis, Alice and Bob must mutually communicate their measurement outcomes using any conventional channel (radio, phone, internet). In order to avoid technical jargon, it is sufficient for the **Instructor** to use AV support (slide or whiteboard) to define the symbol coding used to simulate the correlation analysis. One possible choice of symbol coding is given in Fig. 2b. Whenever Alice and Bob measured different spin projections (colors), the correlation gives a triangle symbol, and if they measured the same spin projection, the correlation gives a circle.

By construction of the demonstration, the correlation analysis of the experiment *with* entanglement gives the same symbol (triangle) for all experimental runs, as shown in Fig. 2a (center data table, right column). On the contrary, in the cross correlation analysis of the experiment *without* entanglement, the two possible correlation symbols can occur, as in Fig. 2a (center data table, left column). Once the cross correlation analysis of the results is done, the **Instructor** introduces the idea of the *Bell test* for entanglement. In our test experience, students at this point already have an intuitive notion of what the effect of entanglement was on the second set of experiments (reduced Bob's possible outcomes). This understanding can be checked with a quick open question to the audience. The simulated quantum entanglement at the particle source effectively *enforced* the correlation analysis to give *always* the same symbol (triangle) for all experimental runs.

The **Instructor** then explains that entanglement between two quantum particles can be operationally defined as a kind of *additional* correlation between the quantum states of the particles that cannot be explained by the laws of classical physics that we experience in our everyday lives. In other words, entanglement is a type of *quantum correlation*. This quantum correlation can be detected in a Bell test by analyzing the joint outcomes of measurements done by independent parties (Alice and Bob) on each particle making an entangled pair. By pointing at the data table with the results of the correlation analysis (Fig. 2a, center data table), the **Instructor** *defines* a simulated Bell test that establishes whether the particles that Alice and Bob each received where entangled or not. If all the correlation results give the same symbol (triangle), the particles were entangled. Otherwise, they were not. Establishing the presence of entanglement in the second set of experiments with this simulated Bell test concludes the demonstration.

**Entanglement and Encryption**

The simulated Bell test described above offers an excellent opportunity to promote student awareness on the connection between fundamental physics and modern technology. After the demonstration is over, instructors are encouraged to immediately follow up with a brief qualitative discussion on the application of quantum entanglement in secure key distribution for message encryption [6].

From previous in-class discussions, students would already be familiar with the idea that in order to transmit a secure message using an encryption algorithm, the emitter (Alice) uses a *key* to convert the secret message text into a string of binary or hexadecimal numbers (encryption), and that the receiver (Bob) uses the *same key* to convert the string back to text (decryption). The message is secure because the encryption/decryption key is only known to Alice and Bob (shared key). A third party (eavesdropper) who manages to learn the shared key, can in principle use it to decrypt the message and learn its content without being noticed. Students thus understand that ensuring the security of the shared key becomes a crucial task in cryptography.

After the demonstration is carried out as described above, the audience is expected to have an intuitive understanding on how different the correlation analysis is, when experiments done by Alice and Bob *with* and *without* entanglement between particles. Building on this understanding, the instructor can use further AV support (slides) to connect the idea of the Bell test with secure key distribution in message encryption as follows:

> Consider the example slides in Fig. 3. The instructor can explain that it is possible to *encode* a shared key in the measurement outcomes of the spin projection measurements carried out by the emitter (Alice) and the receiver (Bob). Alice and Bob must know in advance that the individual particles they each receive from the source belong to an entangled pair. They then carry out spin projection measurements on each particle they receive (pick spheres), and register the measurement outcomes (colors) on their corresponding data tables, as it was done during the demonstration. Alice then uses her set of measurement outcomes to *define* the shared key (i.e., convert colors to binary digits as in Fig. 3). Due to the properties of the entangled state in Eq. (1), Bob learns the encryption key that Alice shared with him by replacing zeros with ones on his measurement outcome records. The secrecy of the distributed quantum key is guaranteed when the joint measurement outcomes pass a Bell test (e.g., the combined result gives all triangles).

Quantum key distribution between Alice and Bob in the presence of an adversary (eavesdropper) is an excellent opportunity to expose students to the practical challenges involved in exploiting quantum entanglement for technological purposes. Instructors at this point can comment on the fragility of quantum entanglement as a resource, in the sense that there are multiple natural and artificial processes that can destroy the quantum correlations introduced by entanglement between quantum particles. In the case of quantum key distribution, an eavesdropper who wants to learn the secret key sent by Alice to Bob can effectively act as an artificial entanglement destruction mechanism. In order to learn the secret key, the eavesdropper can intercept the particle sent to Bob, measure its spin projection, and then put the particle back in the communication channel for Bob to receive it. In the act of measuring the spin projection of Bob's particle, the eavesdropper introduces an irreversible change in wave function of the two particles shared by Alice and Bob, which effectively destroys quantum entanglement. When Bob receives the intercepted particle from the eavesdropper, his measurements no longer correlate perfectly with those of Alice (see Fig. 3, lower panel), and the correlation analysis no longer passes the Bell test. The channel is thus considered insecure and communication is aborted.

We note that the analogy with quantum cryptography proposed in here is not intended to be accurate, as spin (or polarization) measurements are carried out by Alice and Bob along random orientations in real implementations of quantum key distribution. Moreover, several quantum communication protocols do not rely on entanglement to guarantee security [6]. In the proposed teaching experience involving two colored spheres per particle, we can only simulate spin measurements along a single direction in space, with respect to which a particle can be found with its spin either aligned (spin up = red) or anti-aligned (spin down = blue). Despite this unavoidable conceptual limitation of the proposed demonstration, through written evaluations in a preliminary school test that used the slides in Fig. 3, an important percent of students was able to establish the connection between quantum entanglement and cryptography.

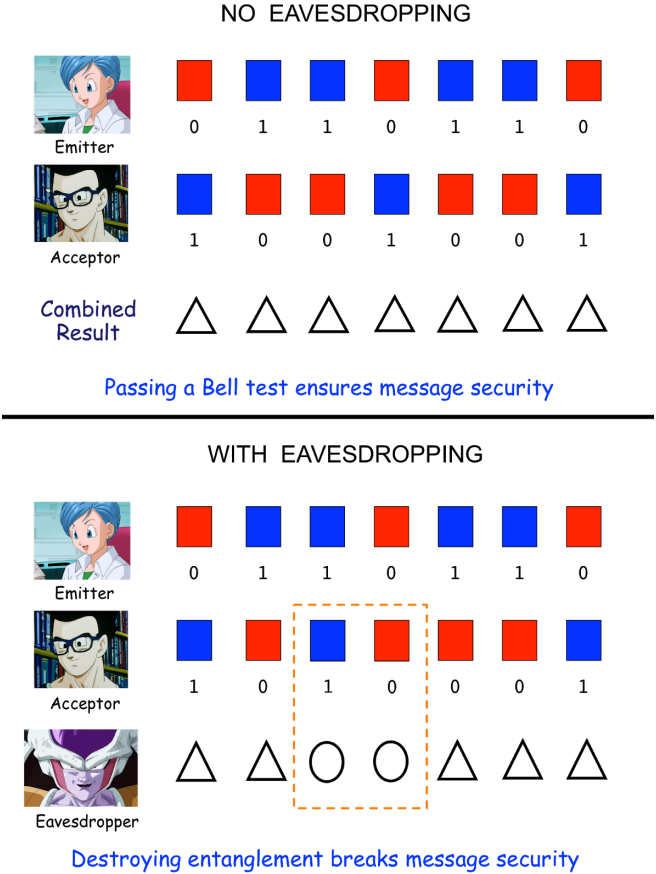

**Figure 3.** Post-demonstration slides illustrating the connection between quantum entanglement and message encryption. (Upper slide) Using a perfect quantum communication channel, Alice and Bob ensure the security of a shared encryption key encoded in their individual measurements outcomes, by establishing entanglement between their particles with a Bell test, i.e., correlation analysis gives all triangles. (Lower slide) In order to learn the secret encryption key that Alice and Bob share, an adversary (eavesdropper) has to measure the spin of one of the messenger particles, which is likely to destroy quantum entanglement in the particle pair, changing the measurement outcomes that Bob (or Alice) can record. The eavesdropper can be detected by noting that the correlation analysis no longer passes the Bell test, i.e., circles can occur.

## Conclusion and Outlook

In this work, we have designed and tested a low-cost interactive live demonstration of quantum entanglement between two particles. The demonstration can be implemented in a regular classroom or in a lecture setting without involving any sophisticated equipment or materials. The demonstration was designed for 9th grade Chilean students, but it could be reproduced in other contexts. The activity aims to explain, in simple and accessible terms, several abstract but technologically useful concepts in quantum mechanics such as projective measurements, indistinguishability of pathways, coherent superpositions, quantum entanglement and Bell tests. The demonstration takes about 20 min to implement, provided that instructors previously discuss about classical message encryption and the quantum bit (*qubit*). We also suggest to use time after the demonstration for explaining the connection between entanglement and secure communication. Our work thus complements previous proposals to introduce quantum physics concepts at the school level [2, 10].

In an effort to increase reproducibility, we comment on the practical need to use particle containers that allow Alice and Bob to pick only one colored foam sphere, representing a spin projection, as randomly as possible. In principle, more than two colored spheres can be put inside particle containers to increase randomness, but this would make the simulation of a projective measurement on the entangled two-particle state very difficult, if not unfeasible. Another practical issue that may arise is the possibility that after completing five repetitions in the set of experiments without entanglement, the data table looks identical to what instructors expect to obtain later in the demonstration, when entanglement between particles is simulated. In this case, instructors should not proceed with the entanglement simulation (second set of experiments), but instead carry out further repetitions of the experiment without entanglement, until the corresponding data table has at least one row with the same color (spin projection) for Alice and Bob. Regarding visual support in a lecture hall setting, slides can be helpful to explain the details of the color coding for the correlation analysis, the concept of a projective measurement on the entangled state, and to introduce the idea of quantum key distribution. However, we stress that as far as visual support is concerned, a classroom whiteboard and colored markers are sufficient tools.

The proposed demonstration can be adapted and implemented in the classroom using either Inquiry-Based Learning (IBL) or Model Based Inquiry (MBI) strategies. Both approaches are suitable mechanisms to allow students develop a deep understanding of abstract concepts [11, 12]. The demonstration described in this work was tested at two Chilean public high schools. In written evaluations, after the demo, students showed a basic understanding of several quantum-related concepts developed in the activity. Evaluations also showed that the proposed demonstration positively impacts the interest of students in physics, and promotes an appreciation on the role of quantum mechanics in modern industry. The validation and analysis of these student evaluations will be the subject of future work.

## Acknowledgements

We thank chemistry teacher Magdalena Loyola and physics teacher Elba Fernandez for permissions to test this demonstration at their workplaces. We also thank Gustavo Lima for discussions on quantum key distribution. The original idea for this work was inspired by a brief comment on the difficulty to explain quantum entanglement to the general public, made by Charles Bennett during a lecture on quantum computing at Universidad de Chile in 2017. F.H. is supported by CONICYT through Proyecto REDES ETAPA INICIAL, Convocatoria 2017 No. REDI 170423, FONDECYT Regular No. 1181743, and The Millennium Scientific Initiative (ICM) through the Millennium Institute for Research in Optics (MIRO).